# A comprehensive review of deep learning in lung cancer

*Farzane Tajidini* [1, *]

[1] *Tabarestan University of Chalus, Chalus, Iran*

**Highlights**

- Deep learning is one of the most promising techniques in medical image processing.
- Deep learning architectures are used extensively in colon cancer analysis.
- Deep learning can be included among the standards in colon cancer diagnosis soon.
- Colon cancer detection and classification can be performed more successfully and faster with CNNs.

## ABSTRACT

To provide the reader with a historical perspective on cancer classification approaches, we first discuss the fundamentals of the area of cancer diagnosis in this article, including the processes of cancer diagnosis and the standard classification methods employed by clinicians. Current methods for cancer diagnosis are deemed ineffective, calling for new and more intelligent approaches. A cancer diagnosis is receiving more attention to define better diagnostic tools, thanks to artificial intelligence. Deep neural networks may be utilized for intelligent picture analysis with effectiveness. This paper presents the fundamental building blocks of machine learning's application to medical imaging—pre-processing, picture segmentation, and post-processing—. In the second section of this paper, all kinds of diseases have been investigated. We give deep learning for each methodology to enable interested readers to test out the described methods on their own diagnostic issues. This manuscript's final section gathers the effectively used deep learning models for various disease kinds. We limit our discussion to skin, lung, brain, and breast cancer due to the length of the paper.

## 1. INTRODUCTION

Cancer is a disease that manifests in many ways and is mostly linked to aberrant cell populations. These cancer cells keep dividing and expanding to become tumors. Lung cancer is cancer that poses the greatest risk to human life globally. Lung cancer is the main cause of mortality globally, claims the World Health Organization [13]. Lung cancer was the cause of 1.37 million deaths worldwide in 2008 [14]. According to the data that is currently available, lung cancer accounts for the majority of new cancer diagnoses worldwide (1,350,000 new cases, or 12.4% of all new cancer cases), as well as the majority of cancer-related fatalities (1,180,000 deaths, or 17.6% of all cancer deaths) [15].

Lung cancer ranked first among causes of death for men and third among causes of death for women in the Global Cancer Observatory database created by the International Agency for Research on Cancer (IARC) in 2018. The database included incidence and mortality rates across 185 countries and 36 types of cancer. Nearly 1.8 million fatalities from cancer were recorded in 2018, accounting for about 18.4% of all cancer-related deaths [1]. Due to the alarming increase in lung cancer fatalities and the disease's excessively high incidence in nature, several cancer control studies and early detection techniques have been developed to reduce mortality.

Typically, lung cancer cure depends upon detecting the disease at the initial stage and effective diagnosing. Effective diagnostic techniques lead to lower incidence rates for lung cancers, and early illness discovery is typically necessary for lung cancer cure. Currently, seven methods can be used to treat lung cancer, including cytology sputum and breath analyses, magnetic resonance imaging (MRI), positron emission tomography (PET), and chest radiographs (CXRs) [2]. The CXR, septum, and CT are radiation-prone, and MRI and PET limit the identification and staging of lung cancer. These techniques also have certain drawbacks.

Additionally, a serum is an intrusive procedure, and its low level of early detection sensitivity and specificity makes it unacceptable. On the other hand, sputum still required further testing due to gene promoter methylation but could identify lung cancer early [3]. Additionally, VOC in urine had excellent sensitivity and specificity but needed a larger sample size [3], whereas CXR has low sensitivity and frequently produces false-negative results [4, 5]. The most reliable method for finding lung cancer nowadays is CT imaging, which provides precise information on the positions and sizes of nodules. Early-stage cancer tumors were found by low-dose CT screening. When compared to conventional radiography procedures, it led to a 20.0% mortality reduction and a noticeably higher percentage of positive screening tests [6].

In this study, we present many CAD systems for detection and classification that have been recently created. Our major goal is to compile important information from many prominent state-of-the-art CNN architectures in tabular style and compare all CNN on various criteria. We also highlight several important critical elements for researchers to remember while identifying and categorizing lung nodules, as well as the recently created models employed in these activities. The sections below make up this review:

A description of the CAD system is presented in Section II. It entails a study of two different CAD systems: a learnt CAD system and a handicraft-based CAD system.

The third section focuses on several cutting-edge CNN architectures, which are the fundamental building blocks of the CAD system. In a tabular style, we highlight and analyze several CNN architectures like LeNet, AlexNet, VGGNet, GoogLeNET, etc. We also provide a few other freshly created models.

We aim to summarize recently proposed CNN-based CAD models in lung nodule detection and classification and their results in section IV. Finally, we conclude this review in section V.

**2. Nodule detection and classification for malignancy estimation**

**2.1. Lung cancer**

By analyzing the CT scan using artificial intelligence techniques, lung cancer CAD aids in the early detection of the disease. These systems sometimes referred to as decision support systems, use the picture segmentation and classification methods depicted in Fig. 3 to evaluate the CT images. By taking into account the form characteristics of nodules, a polygon approximation approach is created to detect nodules. The form and intensity characteristics are collected and fed into the support vector machine (SVM) to find genuine nodules. Therefore, automated picture segmentation is essential to save radiation oncologists from the tiresome contouring labor. Currently, clinical practices frequently employ the atlas-based automated segmentation method. However, the success of this method greatly depends on how closely the segmented image matches the atlas. Deep learning as a component of artificial intelligence is drawing more and more attention in the automated segmentation of medical images in light of substantial advancements in computer vision. In this study, we studied lung cancer-specific automated segmentation methods based on deep learning and compared them to methods based on atlases. The auto-segmentation of OARs with relatively big capacity, such as the heart and lung, performs better than organs with little volume, like the esophagus, at the moment [44].

**2.2. Breast cancer**

Several image processing technologies are being developed in conjunction with artificial intelligence and machine learning to improve the performance of medical and diagnostic operations. For the early diagnosis and categorization of breast cancers, computer-aided detection, or CAD, is a new development in medical imaging. CAD technology helps doctors diagnose patients more accurately while enhancing the quality and sensitivity of medical pictures. With the use of computers, radiologists may quickly find problems. Information on the detection and diagnosis of numerous diseases and anomalies may be found in medical photographs. Radiologists may examine the interior structure using a variety of modalities, and many forms of research have shown much interest in these modalities. Each of these modalities is important in specific medical disciplines [45].

**2.3. Skin cancer**

Melanoma is an aggressive kind of skin cancer that, if left untreated, can be deadly. The previous study mostly focused on creating machine-learning techniques with effective feature descriptors for medical imaging. The performance of CAD systems is constrained by these methodologies, which mostly rely on hand-crafted features. Therefore, DCNNs recently achieved notable medical imaging success, which helped increase performance. One of the most serious types of cancer is skin cancer. Unrepaired DNA breaks in skin cells, which result in genetic flaws or mutations on the skin, are the primary cause of skin cancer. Because skin cancer is more treatable in its early stages and tends to spread to other body areas progressively, it is best identified at an early stage. Given the gravity of these problems, researchers have created a number of early-detection methods for skin cancer. Skin cancer may be detected and distinguished from benign skin cancer and melanoma using lesion criteria, including symmetry, colour, size, form, etc. [46].

**2.4. Lung nodule detection**

To identify lung nodules through screening at an early stage, the lung nodule detection phase is crucial. Most studies on lung nodule detection with a decrease in false positives are currently being done. Researchers can now recognize patches from CT screens without prior knowledge of characteristics thanks to the notable accomplishment of deep learning. Lung cancer early diagnosis is a serious and difficult challenge crucial to a person's survival. The first diagnosis of malignant nodules is often made using chest radiography (X-rays), and computed tomography (CT) scans; however, the potential presence of benign nodules results in incorrect conclusions. The early phases of both benign and malignant nodules exhibit striking similarities. This study's purpose was to summarize the literature on deep learning (ML) algorithms used to detect lung nodules in thoracic CT images to diagnose lung cancer better. The Preferred Reporting Items for Systematic Reviews and Meta-Analyses (PRISMA) standards were followed in constructing this systematic review.

**3. Deep Learning Overview**

Deep learning (DL) is becoming increasingly significant in our daily lives. It has already had a significant influence in fields including voice recognition, self-driving cars, precision medicine, cancer detection, and forecasting. Traditional learning, classification, and pattern recognition methods cannot handle large-scale data sets with their meticulously created feature extractors. The constraints of prior shallow networks that impeded effective training and abstractions of hierarchical representations of multi-dimensional training data were frequently solved by DL, depending on the issue's complexity. Multiple layers of units with highly tuned algorithms and designs make up a deep neural network (DNN).

A machine learning (ML) technology known as a neural network is modeled after and inspired by humans' brains and nervous systems. Processing elements are arranged in input, hidden, and output layers. Each layer's nodes or units are linked to those in the layer below it. Every link has a weight value. At each unit, the inputs are multiplied by the corresponding weights and added [37].

Deep learning has been the most important advancement in computer science in recent years. It has affected almost all scientific disciplines. Businesses and industries are already being disrupted and transformed by it. The top economies and IT firms worldwide are competing to enhance deep learning. Deep learning has already outperformed humans in several fields, such as movie rating prediction, the decision to accept loan applications, the time it takes to deliver a car, etc. [38]. Yoshua Bengio, Geoffrey Hinton, and Yann LeCun, three pioneers of deep learning, received the Turing Award—also known as the "Nobel Prize" of computing—on March 27, 2019 [39]. While significant progress has been made, deep learning still has room for improvement. Deep learning has the potential to enhance human lives through more accurate cancer detection [40], the development of novel medications, and catastrophe prediction [41]. For instance, [42] revealed that a deep learning network could identify at the same level as 21 board-certified dermatologists after learning from 129,450 photos of 2,032 disorders. In grading prostate cancer, Google AI [40] outperformed the typical accuracy of US board-certified general pathologists by 70% to 61%.

**3.1. Convolution Neural Network**

The preferred neural network for computer vision (image recognition) and video recognition is CNN, which is based on the human visual cortex. Other fields, including NLP and drug discovery, also employ it. A CNN comprises a sequence of convolution and subsampling layers, a fully linked layer, and a normalizing (e.g., softmax function) layer, The well-known LeCun et al. [43] 7-layered LeNet-5 CNN architecture for digit recognition. As the sequence of numerous convolution layers moves from the input to the output layers, each layer extracts features with finer resolution.

**3.2. CNN architecture**

Most CNNs are constructed using the fundamental layers presented in Fig. 2. The designs often have several convolutional layers, max-pooling layers, FC layers, and SoftMax layers or output layers as the last layers. LeNet [12], AlexNet [11], and VGGNet [7] are three models from CNN. Other, more effective advanced designs have been put forth, such as the GoogLeNet with Inception units [8], Residual Networks (ResNet), and Densely Connected Convolutional Networks (DenseNet) [10]. These designs share many basic elements, such as convolution and pooling.

However, certain variances are seen in these systems based on connections, computing complexity, and operations carried out at various levels. Because of their cutting-edge performance, AlexNet [12], VGGNet [7], GoogLeNet [8], and DenseNet [10] are typically regarded as the most popular designs. In contrast to the VGG network, which is viewed as a generic architecture, GoogleNet and ResNet are specifically created to process vast amounts of data. Regarding connection, some topologies, like DenseNet [10], are dense. All of the main details pertaining to these cutting-edge architectures are listed in Table 2. The CNN is built layer by layer, as seen below [48-51]:

- **Input layer:**

  Histopathological pictures of breast cancer are entirely loaded into the input layer, producing outputs that feed the first convolutional layer. It is intended to change the size of photos linked to histopathology by subtracting the mean by 256. The pictures added to the layer are three two-dimensional arrays inside RGB channels with an eight-bit depth [48-51].

- **Convolutional layer:**

  The convolutional layer retrieves features by calculating the neuron's output that connects to the input or local regions of the preceding layer. The kernel or filter is the collection of weights that is intertwined with the input. The dimensions of each filter are 3 3, 5 5, or 7 7. Each neuron is also only weakly connected to the area in the previous layer. Stride is used to describing the separation between filter applications. The hyperparameter for the stride is set to two, which is smaller than the filter's size.

  Furthermore, windows that overlap are used by the convolution filter. Additionally, the kernel initializes from the Gaussian distribution with a 0.01 SD. The final layer comprises sixty-four filters initializing from Gaussian distributions and a 0.0001 SD. All amounts of the local weights across ReLU are eventually performed [48-51].

- **Pooling layer:**

  The pooling layer job downsampling feature maps by condensing related feature points into a single point. Receptive field amplification, noise reduction, and dimension reduction are the goals of the pooling layers. The outputs of the pooling layers preserve scale invariance and reduce parameter counts. The last layer employs a mean-pooling strategy with 7 7 receptive fields and a stride of 1 since the relative feature locations are coarse-grained. Further layers use the max-pooling approach with 3 3 receptive fields and a stride of 2 [48–51].

4. **Related Works**

   According to the motive, we separate related work into classification tasks and segmentation tasks in this part. The contributions, processes, and outcomes of each work are then compiled. [16] uses third-party software (LNKnet package), which includes a neural network classifier, to evaluate two suggested texture attributes. Five hundred thirty-six samples are utilized in the experiment for classifier training and 526 samples for testing. Finally, 90% accuracy is attained.

   In [17], to distinguish between benign and malignant breast tumors, Support Vector Machine (SVM), k-Nearest Neighbor (KNN), and Probabilistic Neural Networks (PNN) classifiers are combined with signal-to-noise ratio feature ranking, sequential forward selection-based feature selection, and principal component analysis feature extraction. The use of an SVM classifier yields the highest overall diagnostic accuracy for the detection of breast cancer. On dataset 1 (692 samples of fine-needle aspirates of breast masses), accuracy is at 98.80%, while on dataset 2, accuracy is at 96.33%. (295 microarrays). On datasets 1 and 2, PNN achieves an overall accuracy of 97.23% and 93.39%.

   In [18], a three-layer forward/back ANN classifier and eight characteristics are used to classify four types of H&E-stained breast histopathology pictures. In the experiment, an overall accuracy of almost 95% is attained using 1808 training samples, 387 validation samples, and 387 test samples.

   An automated breast cancer categorization system based on histological pictures is suggested in [19]–[21]. First, features for edges, textures, and intensities are extracted. An ANN classifier is then created based on each of the retrieved characteristics. Thirdly, these classifiers are chosen and aggregated using an ensemble learning technique called "random subspace ensemble" for even higher classification performance. Classification accuracy of 95.22% on a public picture dataset is finally attained.

   Thirty textural characteristics are initially extracted in [22] in order to identify low magnification (10) breast cancer histopathology pictures (H&E stained) into three malignancy classifications. Then, feature selection is used to mine the retrieved features for more useful data. Finally, a PNN classifier is constructed using the chosen features. The experiment's last 65 photos are examined, and overall accuracy of 87% is attained.

   In [23], 70 histopathology photos are randomly chosen as the dataset, and morphological characteristics are retrieved to categorize malignant and non-cancerous cells in histological images. A multi-layer perceptron built on a feed-forward artificial neural network used in the experiment obtains, in that order, 80% accuracy, 82.9% sensitivity, and 89.2% AUC.

   A multiple instance learning architecture for CNN is put out in [24]. Without inter-patch overlap or complete slide coverage, a new pooling layer is suggested that would assist in gathering the majority of the useful characteristics from the patches that make up the whole slide. At 40-, 100-, 200-, and 400-times magnification in the experiment, the accuracy is 89.52%, 89.06%, 88.84%, and 87.67%, respectively.

[25] proposes a novel model that ignores the image's magnification factor for the automated categorization of breast cancer tissue in histological pictures using DCNN. The BreaKHis experimental findings have an 85.3% average accuracy.

In [26], CNN-based features are applied to three-dimensionality reduction techniques such as PCA, Gaussian Random Projection (GRP), and Correlation-based Feature Selection (CBFS) to categorize histological pictures of breast cancer. The BreaKHis dataset, the Epistroma dataset, and the Multi-class Kather's dataset are tested in the experiment. Finally, BreaKHis dataset accuracy is 87.0%, 85.2%, 85.0%, and 81.3%, respectively, at magnifications of 40, 100, 200, and 400. The accuracy on the Epistroma dataset is 94.7%. An accuracy of 84.0% is attained on the Multi-class Kather dataset.

[27] uses transfer learning approaches to train Inception-V3 and Inception-ResNet-V2 for binary and multi-class categorization of breast cancer histopathology pictures. The findings demonstrate that the Inception-ResNet-V2 network performs optimally at a 40x magnification: The picture level accuracy in the binary classification job is 97.90%, and in the multi-classification task, it is 92.07%.

A network for classifying breast cancer histology images (BHCNet) is created in [28]. Three SE-ResNet [81] blocks, one fully connected layer, and one plain convolutional layer are all included in BHCNet. Known as BHCNet-N, each SE-ResNet block comprises N little SE-ResNet modules. The BHCNet-3 achieves an accuracy between 98.87% and 99.34% for the binary classification, and the BHCNet-6 achieves an accuracy between 90.66% and 93.81% for the multi-class classification.

Using a small training dataset, deep learning, transfer learning, and the Generative Antagonistic Network (GAN) is integrated [29] to increase the classification accuracy of breast cancer. First, features are extracted using the improved VGG-16 and VGG-19, then submitted to CNN for classification. Two GAN models—StyleGAN [83] and Pix2Pix [84]—are used to produce 4,800 and 2,912 false pictures, respectively. In the experiment, the BreaKHis dataset and two BreaKHis by GAN-produced datasets are used to test the suggested approach. The trials demonstrate how GAN pictures increased noise levels and impacted classification precision. Finally, the BreaKHis dataset yields the best result. The binary categorization is 98.1% accurate.

In [30] introduces a Principal Component Analysis Network (PCANet) to categorize pictures of Usual Ductal Hyperplasia (UDH) and Ductal Carcinoma In-Situ (DCIS). This study evaluated a dataset of 20 DCIS and 31 UDH pictures, and 10,000 patches were randomly chosen to serve as the model training data. Finally, a 79% accuracy is attained.

A CNN-based categorization approach for the WSIs of breast tissue is suggested in [31]. Two CNNs are trained as part of this effort. The CNN-I divides the WSI into the fat, stroma, and epithelium. CNN-II identifies the stromal areas as either normal or cancer-associated stroma after operating on the stromal regions produced by CNN categorization. I's 646 slices of breast tissue stained with H&E are included in the collection. Two hundred seventy photographs are used in the experiment for training, 80 copies for validation, and 296 images for testing. Finally, a receiver operating characteristic (ROC) area under 0.921 is attained.

In [32], a deep learning algorithm with hierarchical loss and global pooling is developed to distinguish four breast cancer kinds in histopathology pictures. This study examines a dataset of 400 photos, and VGG-16 and VGG-19 networks are used as fundamental deep-learning structures. Two hundred eighty photographs are utilized in the experiment for training, 60 images are used for validation, and 60 images are used for testing. Finally, a 92% accuracy on average is attained.

The classification of five diagnostic breast cancer styles in the histopathological picture is done in [33]. A saliency detector first conducts multi-scale localization of the regions of interest in the diagnostically significant pictures. Next, CNN categorizes the picture patches as five different forms of cancer. The classification maps are then combined for the final categorization. A final accuracy of 55% is attained when 240 photos are utilized in the experiment to test the usefulness of the suggested strategy. The work's defining feature is the involvement of 45 pathologists in the test pictures' final evaluation, which results in an average accuracy of about 65%. Hence, the performance of the proposed method is comparable to the performance of the pathologists that practice breast pathology in their daily routines.

In [34], a strategy for classifying tumor grade, ER status, PAM50 intrinsic subtype, histological subtype, and recurrence risk score using image analysis is created (ROR-PT). Five hundred seventy-one breast tumor instances are utilized in the experiment for training, and 288 are used for testing. It can also discriminate between low-intermediate and high tumor grades (82% accuracy), ER status (84% accuracy), base-like and non-base-like (77% accuracy), lobules vs. ductal (94% accuracy), and high vs. low-medium ROR-PT score (75% accuaracy), among other variables.

In [35], features from pictures are extracted using pre-trained CNN architectures (GoogLeNet, VGGNet, and ResNet), and these features are then input to a fully connected layer. The average pool categorization is used to differentiate between normal and cancerous cells. Two breast microscopic image datasets were used: one was created locally at LRH hospital in Peshawar, Pakistan, and the other is a standard benchmark dataset [51]. Two thousand photographs are utilized for testing in the experiment, whereas 6000 images are used to train the architecture. Finally, a classification accuracy of 97.53% on average is attained.

[36] introduces the Human Epidermal Growth Factor Receptor 2 (HER2) Scoring Contest. The job of automated HER2 scoring has significant clinical importance since HER2 is a significant predictive factor for breast cancer. According to the study, 14 teams submitted 18 papers for review. Eight of the top ten teams employed CNN-based learning techniques in the comprehensive results of all submitted automated methods. It is clear that HER2 automated scoring tasks greatly benefit from CNN-based learning techniques.

5. Discussion

Machine learning and deep learning [48] has been successfully utilized in medical image and healthcare [49] analysis like whole-slide pathology [50], X-ray [51], diabetes [52, 53], breast cancer [54], heart [55], time series [56], Medicinal Plants [57], stock market [58], Stroke [59], etc. Although numerous research has been conducted with the same objectives as those outlined in this paper, it is possible to conclude that there is presently no conclusive approach that can be regarded as effective in every element of lung nodule identification. The parenchyma extraction stage, correct nodule segmentation stage, such as segmentation of just vascular nodules, and generally in the false-positive reduction step are failure points for various techniques. In addition, nodule characterization is crucial for cancer research. Different nodule characterization methods may be suggested based on the growth rate, which may eventually aid in the early identification of cancer and its diagnosis. Another crucial discovery for a patient is the automatic examination of the lung cancer stage. Unfortunately, not many studies with appreciable success rates go this way [47].

6. Conclusion and Future Work

The approaches of breast cancer histopathology image analysis based on artificial neural networks, divided into the conventional artificial neural network and deep neural network methods, are fully discussed in this study. The associated work is organized according to the relevant datasets while summarizing the deep neural network approach.